A DIFFUSIONLESS TRANSFORMATION PATH RELATING $Th_3P_4$ AND SPINEL STRUCTURE

OPPORTUNITIES TO SYNTHESIZE CERAMIC

MATERIALS AT HIGH PRESSURES

by

ARINDOM GOSWAMI

Presented to the Faculty of the Graduate School of

The University of Texas at Arlington in Partial Fulfillment

of the Requirements

for the Degree of

MASTER OF SCIENCE IN CHEMISTRY

THE UNIVERSITY OF TEXAS AT ARLINGTON

AUGUST 2011

ACKNOWLEDGEMENTS

There are so many people I am thankful to without whom this thesis would not be possible. First and foremost, I am thankful to my supervising professor Dr. Peter Kroll for taking me under his wings and constantly motivating and encouraging me to thrive towards my goal, and also for taking the time to critically evaluate this manuscript.

I wish to thank my former academic advisor Dr. Zoltan Schelly, and special thanks to Dr. Krishnan Rajeswar and Dr. Richard Timmons for their interest in my research and for taking time to serve in my dissertation committee.

I wish to thank all my lab-mates Babak Kouchmesky, Munuve Mwania, Ibukun Olubanjo and Nelli Klinova for their keen interest in all my work and for helping me with their inputs/comments and constructive criticism. I wish also to thank my roommates and all the friends from UTA and UT southwestern for their support and encouragement. I am grateful to all the teachers who taught me during the years I spent in school, first in India, and finally in the Unites States.

Last but not the least, I would like to express my deep gratitude to my family who have encouraged and inspired me and supported me throughout my undergraduate and graduate studies. I am also extremely grateful to them for their sacrifice, patience and for being there for me during all my highs and lows.

July 28, 2011

ABSTRACT

A DIFFUSIONLESS TRANSFORMATION PATH RELATING $Th_3P_4$ AND SPINEL

STRUCTURE:OPPORTUNITIES TO SYNTHESIZE METASTABLE

CERAMIC MATERIALS AT HIGH PRESSURES

Arindom Goswami, M.S.

The University of Texas at Arlington, 2011

Supervising Professor:  Peter Kroll

This thesis investigates a transformation path between the $Th_3P_4$-type and the spinel-type structures of hafnium (IV) and titanium (IV) nitride, $M_3N_4$ (M = Hf, Ti) with computational methods. For both compounds, the $Th_3P_4$-type was synthesized experimentally at high-pressure conditions and was quenched to ambient conditions. Computations reveal that while at high pressure $Th_3P_4$-type structures are favored by enthalpy, at low pressure the $Th_3P_4$-type is only meta-stable. The spinel-type is energetically more favorable at zero pressure than the $Th_3P_4$-type in both systems. However, even the spinel-type is surpassed by another ground state modification for both $Hf_3N_4$ and $Ti_3N_4$.

The research presented in this thesis then addresses (i) thermal stability of the $Th_3P_4$ polymorphs and (ii) the conditions necessary to synthesize spinel-type structures of both compounds. The foundation for the study is a simple structural relation between $Th_3P_4$-type and the spinel-type structures: while the $Th_3P_4$-type is based on a body-centered cubic arrangement of anions with cations in interstitial sites, the spinel-type is best described by a face-centered cubic arrangement of anions with cations in interstitial sites. With anions following the Bain path of the bcc-fcc transformation, and cations undergoing a minor shuffle distortion, the $Th_3P_4$ to spinel transformation parallels the martensitic transformation known in many metal systems – only with additional cations in the structure.

Following the Bain correspondence, transformation paths between $Th_3P_4$-type and the spinel-type structures of $Hf_3N_4$ and $Ti_3N_4$ are constructed at different pressures. Along the proposed reaction coordinates total energies in both systems are computed. The obtained energy profiles yield activation barriers and, using an Arrhenius approach, corresponding temperatures. Assuming that the diffusionless $Th_3P_4$ to spinel path will yield the lowest activation barrier for a transformation process of the $Th_3P_4$-type, we can hypothesize on a rate of decomposition of the $Th_3P_4$-type at zero pressure and on a rate of transformation from $Th_3P_4$-type to spinel-type at high pressure.

From the results it is estimated that a thermal stability of the $Th_3P_4$-type of hafnium (IV) nitride is limited to 1200 K at zero pressure. The $Th_3P_4$-type of titanium (IV) nitride exhibits lower thermal stability, the estimation is 950 K. With increasing pressure, both system show increasing barrier height for the transformation process. Accordingly, at the pressure at which $Th_3P_4$-type and spinel-type are in equilibrium – as judged by equal enthalpy – temperatures needed to convert $Th_3P_4$-type into spinel-type are higher. The calculations lead to the proposal that a hitherto unknown spinel modification of $Hf_3N_4$ can be realized by first synthesizing the $Th_3P_4$-type at high pressure, then quenching the system to 3.9 GPa, and at this pressure activating the $Th_3P_4$ to spinel conversion by heating to 1300 K. A similar proposal is made for $Ti_3N_4$. Here a spinel-modification can be synthesized from the $Th_3P_4$-type at 3.9 GPa at a temperature of 1000 K.

TABLE OF CONTENTS

LIST OF ILLUSTRATIONS

LIST OF TABLES

CHAPTER 1

INTRODUCTION

1.1 High Pressure Chemistry of Novel Compounds

Under high-pressure conditions, chemical reactivity of atoms and molecules increases manifold, significant shift in chemical equilibria observed and reaction rates get affected. On compression of any substance, distances between constituting atoms decrease and its electronic structure deforms. This leads to increase of the internal energy of the substance, and formation of denser structures could become energetically favorable on further compression [KOS01], [YAO08]. Shorter interatomic distances and/or higher coordination numbers of the constituent atoms characterize such structures when compared with the lower pressure phases. At high-pressure conditions, formation of compounds with elements in unusual oxidation states can take place. In many cases, these high-pressure phases can be quenched to ambient conditions where they may persist metastably due to slow kinetics of a reverse transformation [BOL09]. Such metastable products can have a number of unique physical and chemical properties, from high hardness and corrosion stability to interesting optoelectronic, magnetic or superconducting properties. Therefore, they can be of potential interest for various industrial applications.

1.2 Nitrides

Nitrides have found extensive use in industry because of their unique physical, chemical, electrical, optical or mechanical properties. The high pressure cubic phase of boron nitride, c-BN is the second hardest material after diamond. AlN, GaN and InN all have interesting semiconductor properties, and each one of them possesses a NaCl-type high pressure phase. Nitrides are used for different applications as refractory ceramics (hexagonal BN, α- and β-$Si_3N_4$, AlN, TiN), hard grinding materials (cubic BN, α- and β-$Si_3N_4$), in light emitting diodes (GaN, InN), as catalysts ($Mo_2N$, $W_2N$), sintering additives, and as electrolyte in lithium batteries ($Li_3N$) [DZI09]. Recent progress in the study of nitrides is mainly due to development of new synthetic routes, which allows large production of compounds and also leads to the discovery of novel compounds [MON07].

Discovery of a new family of high-pressure group 14 cubic spinel nitrides, $γ-A_3N_4$ (A=Si, Ge or Sn) [ZER99] is partially responsible for the upsurge in the interest in the high-pressure nitrides during the last decade. $γ-Si_3N_4$, in particular, has unique combination of hardness and stability (both thermal as well as oxidation) and also a wide band gap semiconductor, which can be used in fabrication of blue LEDs.

Another particular family of ceramics are nitrides of the group 4 elements. High-pressure and high-temperature experiments with transition metals of group 4 resulted in the discovery of exotic family of high-pressure nitrides, $c-Zr_3N_4$ and $c-Hf_3N_4$, with cubic Th3P4-type structure [ZER03].

Since this thesis deals and investigates structural relations among high pressure phases of group 4 elements, we briefly discuss synthesis methods, structure, and experimental and theoretical studies of recently discovered high-pressure zirconium (IV)- and hafnium (IV) nitrides in the following section.

1.2.1 Group 4 transition metal nitrides

At ambient pressure conditions, mononitrides of group 4 transition metals having a cubic rocksalt structure type exist. These mononitrides can also be specified as δ-MN (M= Ti, Zr or Hf). They are used as refractory materials, and as coatings over tools as wear resistant layers. Although they are brittle and hard in comparison to other transition metals, they still exhibit metallic character. All the three mononitrides show superconducting properties with relatively high critical temperatures of 5.8, 10.5 and 6.9 K for TiN, ZrN and HfN, respectively [LEN00]. One more interesting property of these nitrides is their defect structure. They can form several non-stoichiometric structures by creating vacancies on the corresponding lattice sites while still maintaining NaCl-type structure. As for example, in $\delta$-HfN$_x$, x can have values from 0.74 to 1.7 and in $\delta$-ZrN$_x$, x can vary from 0.54 to 1.35 [KRA98]. Presence of vacancies affect their properties significantly. When x > 1, $\delta$-HfN$_x$ and $\delta$-ZrN$_x$ can transform to insulating phases from metallic phases with an increase in nitrogen content. $\delta$-TiN$_x$ maintains its metallic character for all the known compositions [AND97].

Out of all the stoichiometric M$_3$N$_4$ compounds of group 4 transition metals, only orthorhombic zirconium (IV) nitride, o-Zr$_3$N$_4$ was known for long [LER96]. The o-Zr$_3$N$_4$ was found to be insulating and diamagnetic. It decomposes above 1100 K on heating in presence of nitrogen yielding $\delta$-ZrN and N$_2$. It oxidizes in air at 800 K [LER97]. The high-pressure investigation of group 4 transition metal nitrides of the type A$_3$N$_4$ was an extension of study of spinel nitrides. Discovery of novel $\gamma$-A$_3$N$_4$ (A=Si, Ge or Sn) has set the trend of investigation of possible existence of stable spinel nitrides of similar tetravalent elements like group 4 transition metals. On performing experiment at high temperature and pressure, treating Ti, Zr and Hf (or their mononitrides) with molecular nitrogen in a LH-DAC, Zr$_3$N$_4$ and Hf$_3$N$_4$ with cubic Th$_3$P$_4$ structure type results (for Zr-N system at 15.8-18 GPa and 2500-3000 K and for Hf-N system at 18 GPa and 2800 K) [ZER03]. However, for Ti-N systems no other phase besides $\delta$-TiN was observed in the pressure range 17.5-25 GPa and temperature ranging from 1500 to 3000 K. Examining the structure and surface morphology using XRD and EDX proved that the resulting structures are Zr$_3$N$_4$ and Hf$_3$N$_4$ with cubic Th$_3$P$_4$ structure type [ZER03]. The cubic Th$_3$P$_4$ structure type has cations eight-fold coordinated by nitrogen atoms, while nitrogens are six-fold coordinated by metal atoms (discussed in chapter 3). As the cation coordination number is higher (eight-fold) than NaCl-type mononitrides and o-Zr$_3$N$_4$ (six-fold) as well as in the hypothetical spinel nitrides (four- and six-fold), the density of c-Zr$_3$N$_4$ is 13% higher than o-Zr$_3$N$_4$. c-Zr$_3$N$_4$ and o-Hf$_3$N$_4$ are binary nitrides, first of its kind, with such high coordination number [ZER03]. Theoretical studies on their electronic properties followed after, which also supported structural assignment and low compressibility of c-M$_3$N$_4$ [KRO03b] as suggested by experiments. Theoretical work on nitrogen rich part of Ti-N, Zr-N and Hf-N P-T phase diagrams provides equilibria conditions between δ-MN + $N_2$ and c-$M_3N_4$ phases [KRO04]. Interest on these materials increased rapidly with the recent finding that thin films made of c-$Zr_3N_4$ are significantly harder than those of δ-ZrN. Also, it functions as a better wear resistance tool in low-carbon steel manufacturing machine compared to δ-TiN [CHH05], which is a well-known wear resistant material. To sum up, the recently discovered cubic $Zr_3N_4$ and $Hf_3N_4$ are of utmost interest for use in industry for their microelectronics and wear-resistant applications.

### 1.3 Goal of this work

This work exploits a structural relation between the $Th_3P_4$-type, the structure type of novel high-pressure $Hf_3N_4$ and $Zr_3N_4$ and the spinel type. We will construct a transition path from $Th_3P_4$ to spinel phase. While the spinel type has not been observed among the group 4 transition metal nitrides, it is the target of this work to investigate under which conditions it may form from the existing $Th_3P_4$ –type. We, thus, will follow in two directions, which mark the goals of this work.

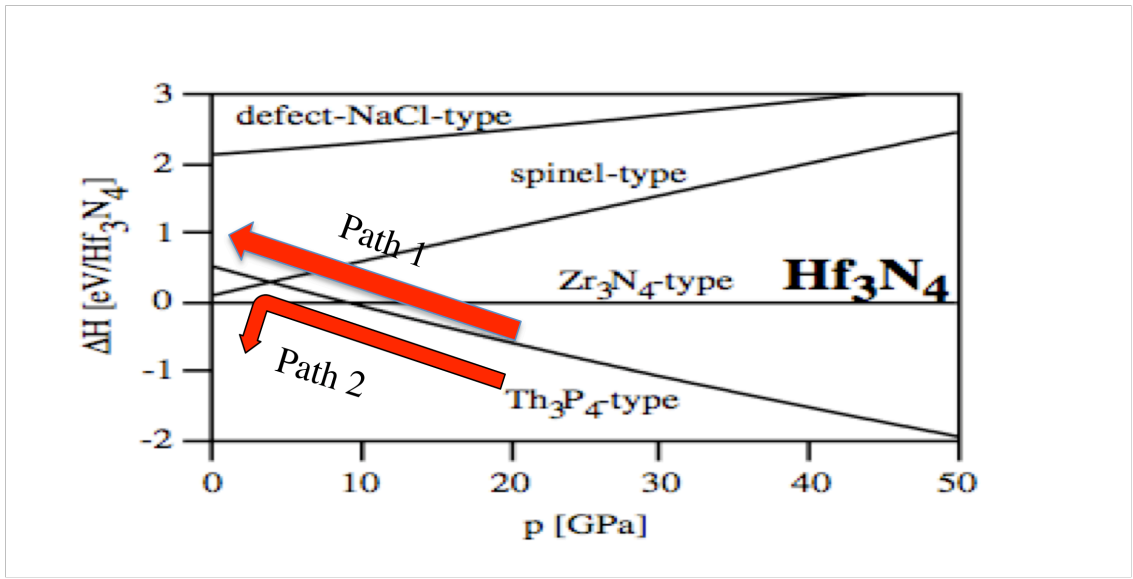

Figure 1: Goal of this work

First, we will compute the energy along the path from $Th_3P_4$ to spinel at ambient pressure. The resulting energy profile with its activation barrier will yield information about the thermal stability of the $Th_3P_4$ type at ambient pressure. Second, we will repeat these calculations along the path at the transition pressure of the $Th_3P_4$-spinel transition. Since the enthalpy of both structures at this pressure is equivalent, a transformation will not yield excess energy. Therefore, the activation barrier will yield information about the temperature needed to activate the $Th_3P_4$-spinel transition.

CHAPTER 2

METHODOLOGY

2.1 Density Functional Theory (DFT)

Electrons and atomic nuclei constitute any solid material. The laws of quantum mechanics govern their properties. In principle, physical properties in a solid can be calculated by solving the many-body Schrodinger equation. Density functional theory (DFT) is the quantum mechanical approach used to solve the many-body Schrodinger equation. Electron density is the fundamental quantity in DFT. Electronic density replaces the complicated many-electron wave function, $\Psi$ for efficient calculation of the Schrodinger equation, made possible by DFT. Density functional theory (DFT) was introduced in 1960's by Hohenberg and Kohn [HOH64] and Kohn and Sham [KOH65]. The major breakthrough that came out from these two approaches are: a) ground-state energy can be written as a unique functional of the electron density [Hohenberg-Kohn] and b) introduction of an energy functional for the total energy of the electronic system to derive Schrodinger-like single-particle equations [Kohn-Sham]. DFT is very effective in computing electronic structure of materials. DFT is used to make predictions of several material properties such as structure, vibrational frequencies, elasticity coefficients, electronic and magnetic properties, paths in catalytic reactions, etc [CUE03]. Out of the many programs available for solid state calculations we have chosen the VASP (Vienna Ab-initio Simulation Package)-code, since it allows efficient and fast calculations.

*2.1.1. VASP* (Vienna Ab-initio Simulation Package)

VASP is used to perform first-principles electronic calculations. VASP implements the DFT combining a plane-wave basis set with the total energy pseudopotential approach. The pseudopotentials used are based on the Projector-Augmented-Wave (PAW) method [KRE1999]. The PAW potential, in principle, is an all– electron potential and therefore allows the calculation all-electron properties.

The Generalized-Gradient Approximation (GGA) is used to treat the exchange-correlation energy of the electrons. Preferring the GGA over the Local Density Approximation (LDA) is based on the fact that gradient corrections yield more accurate results for the energy when comparing atoms in different environment. GGA reproduces experimental transition pressures while LDA artificially favors higher coordinated structures. All our results are obtained from well-converged structures with respect to cutoff energy (500 eV for nitrides and oxides) and k-point sampling.

## 2.2 Computational approach

While searching for new high-pressure modifications, the approach involves computations for structures from crystal database library. In a first step, a selected structure type is optimized in atomic positions and lattice parameters. Comparing the energies of various candidate structures the lowest energy modification for a compound is determined and, its energy and volume can be compared with all the other structure types [HOR06]. In a second step, the energy E is calculated for a series of volumes around the minimum energy structure for each candidate. The resulting energy-volume (E-V) data is used to calculate the enthalpy-pressure state function. The pressure p can be extracted from the E-V data by numerical differentiation. Using $p=-\delta E/\delta V$ the enthalpy $H= E + pV$ is obtained. Though it is the Gibbs energy, which determines thermodynamic stability, we neglect entropy contribution $-T\Delta S$ to the enthalpy. This approach is justified for compact compounds without sources of significant disorder, since entropy contributions $–T\Delta S$ are typically small in comparison to changes in relative enthalpy $\Delta H$ within a few GPa of pressure. Thus, comparing the enthalpy-pressure state functions for different structure types provides the structure with lowest enthalpy at a given p. The best way to extract the transformation pressure is to plot the enthalpy difference with reference to a reference state, usually the ground state.

CHAPTER 3

INTRODUCTION TO THE TRANSFORMATION PATH

3.1 The Bain Correspondence

bcc and fcc structures are related to each other by a tetragonal strain, either compression or extension. The mechanism governing the relation between these two structures is known as the Bain Correspondence. The bcc structure has a c/a ratio equal to 1. There is compression along the z-axis and a uniform expansion along the x and y axes resulting in change in c/a ratio to √2 in the resulting fcc structure [KRO03], [KIB07].

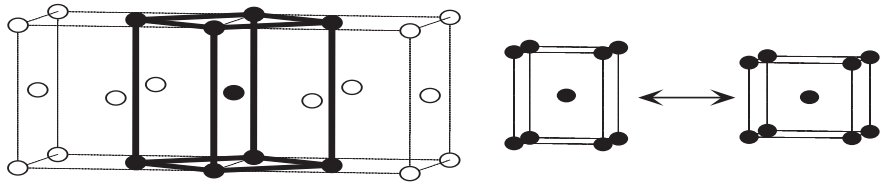

Figure 2: (Left) two unit fcc cells with a smaller body centered tetragonal cell marked in the center. By a homogeneous compression in the direction of arrows and an expansion in the plane normal to it, the bcc structure (right) is obtained.

No shuffle is required in this transformation, meaning, the atoms remain in their crystallographic positions. The Bain correspondence, which related the fcc to bcc arrangement in metal structures, is now used to connect the anion arrangement in the spinel structure, which is approximately fcc, to the arrangements of anions in the $Th_3P_4$ structure, which is approximately bcc. It will turn out that some cations will undergo additional shuffle. This shuffle, however, is surprisingly small and corresponds only to a change in the coordination environment, but not to a significant diffusion. The transformation from spinel to $Th_3P_4$, or vice versa, will proceed through an energy barrier with height $\Delta E_a$. Similar to conventional chemical kinetics, the rate of transformation can now be described by an Arrhenius-type law. While the barrier must be overcome, the activation energy $\Delta E_a$ will correspond to a specific temperature. According to $T_a = \Delta E_a /K$ (or $T_a = \Delta E_a /R$, if one mole of particles is referenced). $T_a$ is about the temperature at which the rate becomes significant. Depending on the direction of the process and the overall reaction enthalpy, $T_a$ can be interpreted differently. If the reaction proceeds from a state with higher enthalpy to a state with lower enthalpy, $T_a$ provides a measure for the (maximum) thermal stability of the initial state. If the initial and final state are equal in enthalpy, $T_a$ may indicate the temperature above which equilibrium can be reached within a reasonable time.

3.2 Description of the structures: The spinel structure (fcc) and its bcc counterpart Th3P4 type

The spinel structure ($A_2BX_4$) adopts space group symmetry Fd-3m. In spinel the cations exist in two different environments: cations (A) in ½ of the octahedral sites and cations (B) in 1/8 of the tetrahedral sites. Anions are located in position 32e (x,x,x) with the free positional parameter $x = 3/8 + x_{sp}$. If $x_{sp}= 0$, the anions are at the points of a face centered cubic lattice (fcc). For all spinel structures the parameter $x_{sp}$ is very close to zero. Note that the spinel structure has two parameters only: the lattice parameter and the positional parameter of the anions.

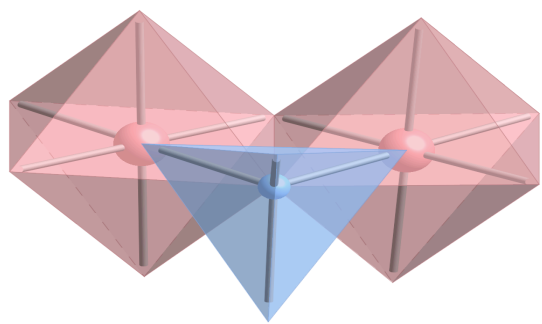

Figure 3: Spinel structure

The $Th_3P_4$ structure ($A_3X_4$) adopts space group symmetry I-43d [OKE96]. Cations (A) fill sites in 12a and anions (X) fill sites in 16c ($x_b$, $x_b$, $x_b$). Similar to the spinel structure, the $Th_3P_4$ structure has only two parameters: the lattice parameter and the positional parameter of the anions. With $x_b=0$, the anion structure becomes ideal bcc. In the $Th_3P_4$-type, $x_b$ is close to -1/12. In this case, the coordination polyhedron around each cation is a bisdisphenoid. This 8-fold coordination is composed of two interpenetrating tetrahedral, one stretched and the other one flattened. Anions are coordinated 6-fold in $Th_3P_4$ and the environment is described by a metaprism, which is between a trigonal prism and a trigonal anti-prism which ideally equals an octahedron.

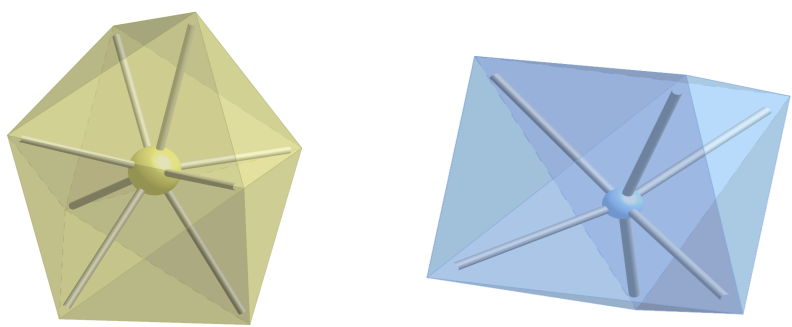

Figure 4: $Th_3P_4$ structure. (left) cation position, {A}$X_8$-bisdisphenoids and (right) anion position, {X}$A_6$-metaprisms

### 3.3 Crystallographic description of the transformation path

As pointed out before, spinel and $Th_3P_4$ structures adopt Fd-3m and I-43d space group symmetries respectively. To describe a path between the two structures, we choose the sub-group/ super-group approach. A common sub-group for both structures is I-42d. Description of these two structures using subgroup I-42d is given below:

Table 1: Crystallographic description of the transformation path

| | Spinel | | | | Th$_3$P$_4$ | | |
| | a=a$_{sp}$, c/a=√2 | | | | a=c=a$_b$ | | |
| Atom | Wy | x | y | z | x | y | z |
| --- | --- | --- | --- | --- | --- | --- | --- |
| A | 4a | 0 | 0 | 0 | 0 | 0 | 0 |
| B | 8d | ½ | ¼ | 1/8 | 5/8 | ¼ | 1/8 |
| X | 16e | 0 | ¼-2x$_{sp}$ | 3/8+ x$_{sp}$ | x$_b$ | ¼+2x$_b$ | 3/8+ x$_b$ |

Here, a$_{sp}$ and a$_b$ are lattice parameters of spinel and Th$_3$P$_4$ respectively, and x$_{sp}$ and x$_b$ are the anion positional parameter of the conventional settings of spinel and Th$_3$P$_4$ structures, respectively. Wy stands for Wyckoff position.

Inspecting the table above, one observes that the cations in A position remain in their sites. Cation in B position change only in their x-coordinate. Anions change x, y, and z. However, given the fact that the anion structure parameter of spinel and Th$_3$P$_4$, x$_{sp}$ and x$_b$, are of small values, their move is very limited. Thus, as in the standard Bain transformation, the dominant change in structure is made by varying the lattice parameter a and changing the ratio c/a. Overall, we have 6 parameters to describe the transition. However, exploring the complete 6-dimensional potential energy (or enthalpy) surface is impossible.

As an approximation of the transformation path, a one-dimensional reaction coordinate has been constructed applying linear interpolation of all structural parameters. Static calculation performed on the intermediate structures yield an activation barrier, ΔE$_a$. <u>Next, we did calculations at the transition pressure, which yields the temperature of synthesis for this phase transformation.</u>

CHAPTER 4

RESULTS

4.1 Variation of energy and enthalpy under pressure

All hard materials, including diamond and c-BN, derived at high pressure are metastable at ambient pressure. These materials can withstand high temperatures during compaction treatments necessary to form a tool from the powdery source material as well as during actual performance as a tool. Therefore, the importance of high "metastability" or thermal stability cannot be ignored. The term thermal stability refers to the temperature at which decomposition or back transformation to the low-pressure polymorphs occurs [SCH03]. In order to access thermal stability (our first goal, as stated before), we need to follow a definite regime as stated previously in section 2.2.

First, taking the internal energy E as a function of the volume V for both $Hf_3N_4$ and $Ti_3N_4$ energy-volume (E-V) diagrams are constructed, considering three structure types –spinel, $Th_3P_4$ and lowest energy form o-$Zr_3N_4$ (in case of $Hf_3N_4$) and $CaTi_2O_4$ (for $Ti_3N_4$) in each case. Next, enthalpy difference-pressure (dH-p) diagrams constructed for $Hf_3N_4$ and $Ti_3N_4$ with reference to lowest energy state. The dH-p diagrams for $Hf_3N_4$ and $Ti_3N_4$ shows that, cubic $Th_3P_4$ will be adopted at high pressures, although it will be metastable at 0 GPa. Also from the diagram, the synthesized cubic $Hf_3N_4$ is a metastable compound having appreciable difference in energy (0.5 eV/f.u.) to the lowest energy $Zr_3N_4$-type (not yet synthesized) modification of $Hf_3N_4$. Spinel phase is metastable for any pressure-temperature condition. Similarly, cubic $Ti_3N_4$ (yet to synthesized) is a metastable compound with energy difference ~0.75 eV/f.u. to the lowest energy $CaTi_2O_4$-type ( also not yet synthesized) [KRO04]. Also from the diagrams, the spinel modification is always lower than $Th_3P_4$ in energy but close to the $Zr_3N_4$ type at zero pressure, and is unfavorable towards compression at high pressure. For spinel -> $Th_3P_4$ type phase transitions the transition pressures calculated are 3.9 and 4.9 for $Hf_3N_4$ and $Ti_3N_4$ respectively. Cubic $Hf_3N_4$ was synthesized around 18 GPa and can be quenched to 0 GPa. Thus, it may be possible to attain a phase that is thermodynamically metastable at zero pressure. If we can provide the temperature corresponding to spinel -> $Th_3P_4$ transition pressure (which was 3.9 and 4.87 for $Hf_3N_4$ and $Ti_3N_4$ respectively), it is possible to attain the spinel structure type on quenching. We want to predict an estimate of that temperature in this work.

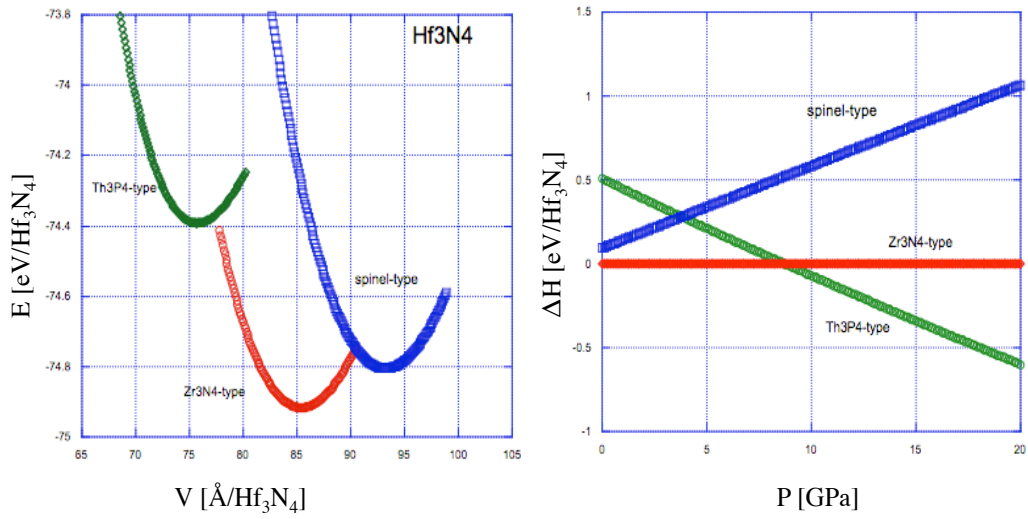

Figure 5: Energy-volume (E-V) (left) and enthalpy-pressure (ΔH-p) diagrams (right) of $Hf_3N_4$.

Table 2: Structure type, energy $E_0$ and volume $V_0$ calculated at zero pressure, relative energy ΔE with respect to the ground state configuration for the three polymorphs of interest of $Hf_3N_4$

| Structure type | $E_0$ | $V_0$ | ΔE |
|---|---|---|---|
| o-$Zr_3N_4$ | -74.908 | 85.28 | 0 |
| $Th_3P_4$ | -74.392 | 75.66 | 0.516 |
| Spinel | -74.806 | 93.15 | 0.102 |

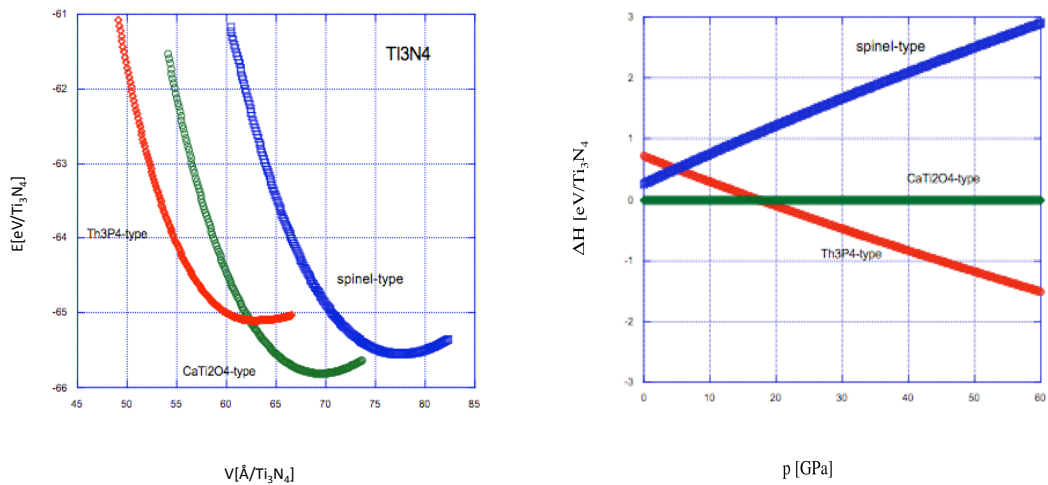

Figure 6: Energy-volume (E-V) (left) and enthalpy-pressure (ΔH-p) diagrams (right) of $Ti_3N_4$.

Table 3: Structure type, energy $E_0$ and volume $V_0$ calculated at zero pressure, relative energy ΔE with respect to the ground state configuration for the three polymorphs of interest of $Ti_3N_4$

| Structure type | $E_0$ | $V_0$ | ΔE |
|---|---|---|---|
| $CaTi_2O_4$ | -65.804 | 69.41 | 0 |
| $Th_3P_4$ | -65.097 | 62.75 | 0.707 |
| Spinel | -65.550 | 77.56 | 0.254 |

### 4.2 Energy profile calculation along the reaction coordinate

A transformation scheme between spinel and $Th_3P_4$ retaining the symmetry of their common subgroup I-42d, was given at table 1 (page 13) before. As it implies, the transformation can be defined within six-dimensional configuration space as there are six parameters, which are changing during the transformation. A reaction coordinate was constructed by simultaneous linear interpolation of all the six structural parameters. Static calculations of these intermediate geometries yield an upper boundary of the activation barrier. The results of these calculations for $Hf_3N_4$ are shown below. The same regime was followed for $Ti_3N_4$ too.

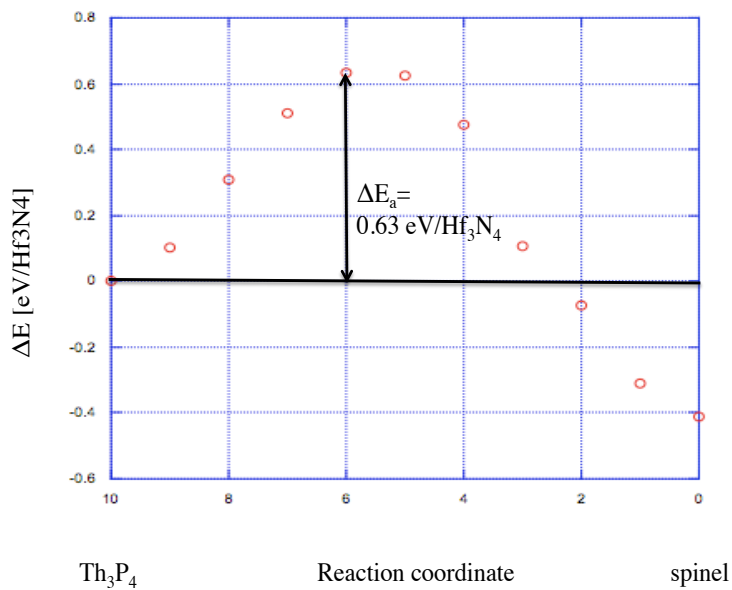

Th$_3$P$_4$          Reaction coordinate          spinel

Figure 7: Energy profile diagram along the reaction coordinate for Hf$_3$N$_4$ (at 0 GPa). The linear interpolation was done for 9 points between Hf$_3$N$_4$ with spinel and Th$_3$P$_4$ structure types. All the six parameters were interpolated simultaneously. The activation energy Ea is about 0.63 eV per Hf$_3$N$_4$, the excess energy is equal to the energy difference between the spinel and the Th$_3$P$_4$ structure types, 0.4 eV per Hf$_3$N$_4$. In the x-axis 0 represents the spinel type while 10 is the Th$_3$P$_4$ type. (same notation is followed in the subsequent graphs).

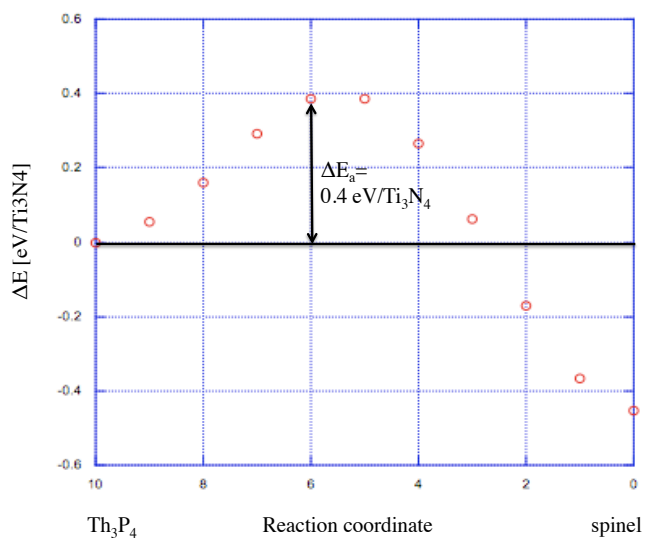

Th$_3$P$_4$          Reaction coordinate          spinel

Figure 8: Energy profile diagram along the reaction coordinate for Ti$_3$N$_4$ (at 0 GPa). (Followed the same procedure as for Hf$_3$N$_4$). Here, the activation energy Ea is about 0.4 eV per Ti$_3$N$_4$, the excess energy is equal to the energy difference between the spinel and the $Th_3P_4$ structure types, 0.48 eV per $Ti_3N_4$.

4.2.1 Results of energy profile calculation

Activation barrier can be calculated along a continuous path following a continuous path of structural transformation [figure 6 and 7]. From these static energy profile calculations at 0 GPa, barriers of 0.63 eV/f.u. and 0.4 eV/f.u. were estimated for the spinel->$Th_3P_4$ transformation for $Hf_3N_4$ and $Ti_3N_4$ respectively. On translating these energies per atom into temperature it gives 1125 K and around 800 K for $Hf_3N_4$ and $Ti_3N_4$ respectively, which corresponds to the thermal barrier. On comparing to typical bond-energies necessary to break during a reconstructive process of Hf-N and Ti-N bonds, these are significantly smaller barriers.

Now to achieve our second goal, that is to estimate the temperature of synthesis of still hypothetical spinel modification from cubic $Th_3P_4$ type on quenching, linear interpolation were done at the transition pressure (same procedure as before but previous calculations were done at 0 GPa). The results of these calculations are shown below:

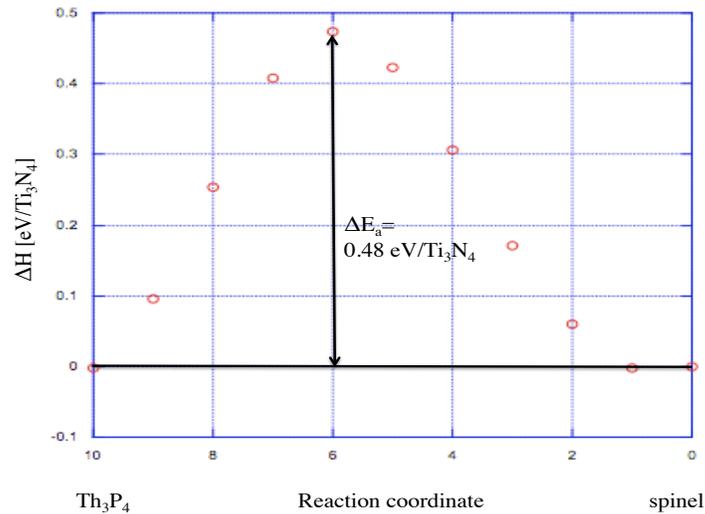

Figure 9: Enthalpy profile diagram along the reaction coordinate for $Ti_3N_4$. Calculations were done at transition pressure ($P_t$=4.9 GPa).

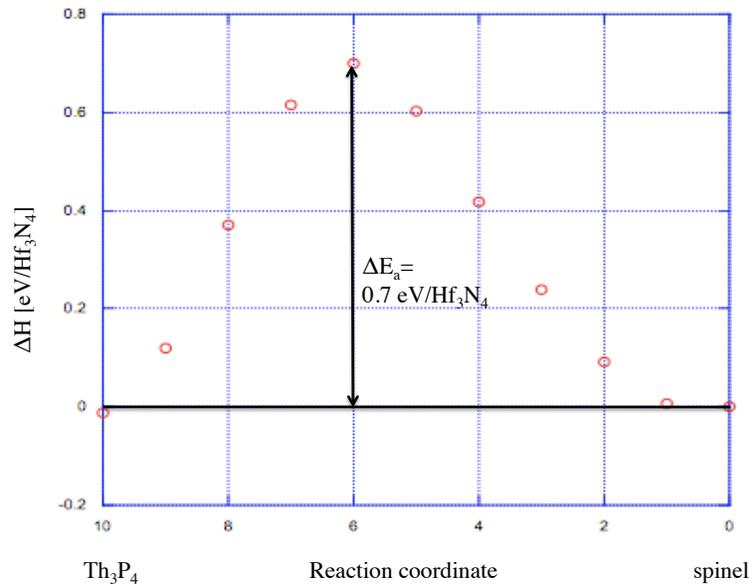

Figure 10: Enthalpy profile diagram along the reaction coordinate for $Hf_3N_4$ ($P_t$=4.9 GPa).

The same formalism was applied as before and calculations were carried out for the transformation at the transition pressure to access the temperature of synthesis. Barriers of 0.7 eV/f.u. and 0.48 eV/f.u. were estimated for $Hf_3N_4$ and $Ti_3N_4$ respectively. These energies corresponds temperatures of around 1200 K and 900 K respectively.

CHAPTER 5

SUMMARY AND CONCLUSION

Using a simple structural relation paralleling martensitic transformations, energy profiles along a transformation path from $Th_3P_4$-type to spinel-type structures have been computed for hafnium (IV) and titanium (IV) nitride, $M_3N_4$ (M = Hf, Ti). At zero pressure, where a spinel modification is energetically more favorable than a $Th_3P_4$ modification, the activation barriers for the $Th_3P_4$ to spinel transformations are 0.63 eV/f.u. and 0.4 eV/f.u. for $Hf_3N_4$ and $Ti_3N_4$, respectively. It is estimated, therefore, that at temperatures of 1120 and 800 K, respectively, this transformation will play a significant role in the decomposition of $Th_3P_4$-type $Hf_3N_4$ and $Ti_3N_4$. The same transformation path is used to predict the conditions necessary to synthesize a (meta-stable) spinel modification from the $Th_3P_4$-type at elevated pressures, when both modifications are assumed to be in equilibrium. A barrier of 0.7 eV/f.u. at 3.9 GPa for $Hf_3N_4$ indicate that annealing of the $Th_3P_4$-type at this pressure may yield a spinel modification. The conditions for synthesizing a spinel modification of $Ti_3N_4$ from a $Th_3P_4$-type are computed from the barrier obtained at 4.9 GPa. 0.48 eV/f.u. correspond to 900 K.

BIOGRAPHICAL INFORMATION

Arindom Goswami was born in Assam, India on 1st of August 1984. He obtained his Bachelor's Degree in Chemistry from Jorhat Science College (now known as Jorhat Institute of Science and Technology) under Dibrugarh University, Assam India in 2006 August. Then, he completed his Master's Degree in Physical Chemistry from University of Delhi in July 2008. After showing immense interest in research, he joined The University of Texas at Arlington in fall 2009 and joined Dr. Peter Kroll's computational laboratory for pursuing research in the fields of nanosciences, ceramics and solid-state chemistry. It is the ultimate aim of his life to contribute to the field of Nanotechnology by doing cutting-edge research.